# Simulation Based: Study and Analysis of Routing Protocol in Vehicular Ad-hoc Network Environment

Mrs. Vaishali D. Khairnar
Research Scholar in Institute of Technology
Nirma University – Ahmedabad

Dr. S. N. Pradhan
Professor in Institute of Technology
Nirma University – Ahmedabad.

Abstract: A Vehicular Ad hoc Network (VANET) consists of vehicles which communicate with each other and exchange data via wireless communication links available between the vehicles which are in communication ranges of vehicles to improve the road safety in city. The communication between vehicles is used to provide road safety, comfort and entertainment. The performance of communication depends on how better routing takes place in the network. Routing data between the source and destination vehicle depends on the routing protocols being used in vehicular ad-hoc network. In this simulation based study we investigated about different ad hoc routing protocols for vehicular ad-hoc network (VANET). The main goal of our study was to identify which ad hoc routing protocol has better performance in highly mobile environment of vehicular ad-hoc network. We have measured the performance of routing protocols using 802.11p in vehicular ad-hoc network in which we considered the scenario of city (i.e. Route between Nerul and Vashi) where we have take 1200 different types of vehicles and checked their performance. Routing protocols are selected after the literature review. The selected protocols are then evaluated through simulation under 802.11p in terms of performance metrics (i.e PDR & E2E delay).

## 1. INTRODUCTION

As the sharp increase of vehicles on roads in the recent years, driving has not stopped from being more challenging and dangerous. Roads are saturated, safety distance and reasonable speeds are hardly respected, and drivers often lack enough attention. Without a clear signal of improvement in the near future, leading car manufacturers decided to jointly work with national government agencies to develop solutions aimed at helping drivers on the roads by anticipating hazardous events or avoiding bad traffic areas. One of the outcomes has been a novel type of wireless access called Wireless Access for Vehicular Environment (WAVE) dedicated to vehicle-to-vehicle and vehicle-to-roadside communications. While the major objective has clearly been to improve the overall safety of vehicular traffic, promising traffic management solutions [1-2].

Equipped with WAVE communication devices, cars units form a highly dynamic network called a Vehicular Ad Hoc Network (VANET), which is a special kind of Mobile Ad-Hoc Networks (MANETs). While safety applications mostly need local broadcast connectivity, developed for intelligent transportation systems (ITS) would benefit from unicast communication over a multi-hop connectivity. Moreover, it is conceivable that applications that deliver contents and disseminate useful information can flourish with the support of multi-hop connectivity in VANETs. Although number of routing protocols has been developed in MANETs, all do not apply well to VANETs. VANETs represent a particularly challenging class of MANETs. They are distributed, self-organizing communication networks formed by moving vehicles, and are thus characterized by very high node mobility and limited degrees of freedom in mobility patterns [1-5].

## 2. OVERVIEW OF IEEE 802.11p PROTOCOL

IEEE 802.11p is applied in the VANET – it is more expensive but able to send data over medium distances and to distribute the information in geographical regions via multi-hop communication fig.1.

IEEE 802.11p is a draft amendment to the IEEE 802.11 standard dedicated to vehicular environments. It is derived from IEEE 802.11a standard, but PHY and MAC layers are modified to support low-latency communication among vehicles. IEEE 802.11p operates at a frequency band specifically allocated for road safety, such as 5.850–5.925 GHz (75 MHZ) in the US and 5.875–5.90 GHz (30 MHz) in Europe with possible future extension, defines data rates from 3 to 27 MHz for 10 MHz channels (optionally 6 to 54 MHz for 20 MHz channels), OFDM modulation and maximum power levels of 44.8dBm. The basic MAC is the same as the well known IEEE 802.11 Distributed Coordination Function (DCF). It adopts concepts from Enhanced Distributed Channel Access (EDCA) of 802.11e, like Access Category (AC) and Arbitrary Inter-Frame Space (AIFS), in order to differentiate priorities among applications.

IEEE 802.11p is designed as a multi-channel scheme, where nodes can switch between channels (US) or transceive on multiple channels simultaneously (dual transceiver in Europe). For advanced networking algorithms, we use standard-compatible extensions to control radio parameters (transmit power and others) on a per-packet basis [13-16].

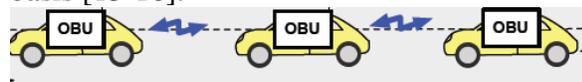

Fig.1 V2V Communication in VANET

# 3. ROUTING PROTOCOLS IN VEHICULAR AD-HOC NETWORKS FOR MESSAGE SCATTERING

Movements of vehicles are described in the form of distinct clusters, which is also an observable scene. Therefore, V2V communication within a single group of vehicles depends on the range of wireless coverage being used; commonly deal with physical aspects of the networks. When it comes to the internetworking, more generic - inter-cluster communication, where source vehicle from a group is unreachable in contact with destination vehicle from other (could be same) group creates the communication discontinuity. To overcome the problem like these terminations, various routing protocols are employed for continuous V2V communication manageability purposes. Hence, the concept of hopping where destinations achieved by mean of relaying on other intermediate node is used specifically.

The various experimentations done have identified that the popular Mobile ad-hoc network (MANET) routing protocol AODV is not much suitable for Vehicular ad-hoc network (VANET) environment. Therefore they have proposed its modified version considering its core parameter- direction - as a focal point for route discovery and named it as DAODV. This proposed algorithm via theoretical simulation results has shown the reduction of overall overhead of route discovery than AODV in comparison of highly performance measures. The results like recovery from broken links and route expiration time are improved with respect to number of hops, nodes, and speed with the projected version. The logical aspect of the intended protocol is only positive while focusing on a particular mobility model for example, Manhattan, in this case. Whereas, there are diversified models available depending on their mobility constrains which need to be plotted for generic acceptance of DAODV.

An application oriented study done by NEC Europe researchers for analyzing the networks mobility and their deployable ability approaches in VANET environment.

The convergence of IP mobility and Network mobility within VANET is clearly highlighted with the term "*VANEMO*". It differentiates the related applications of VANETs by mean of "safety" and "non-safety" grouping with classifications into economical and functional performances, and deploy ability requirements. The two main approaches of connectivity i.e. MANET-centric and NEMO-centric are discussed. With previous approach (which agrees the scope), mobile nodes communicate with each other by utilizing MANET layer for ad hoc routing protocols towards NEMO for high end (infrastructure) connectivity. The latter approach is directly connecting the physical layer of nodes with the NEMO layer. This difference is required to maintain the connectivity tasks of safety (reliable) and non-safety (general ad hoc) applications within different mobility scenarios. Similarly , the application aspect of public safety issues in transportation system and how to mitigate the cases of accident (through chain collision) cases by using vehicle to vehicle communication paradigm are discussed. The major concepts discussed are of Cooperative Collision Avoidance (CCA) with the help of Wireless Collision Warning Messages (W-CWM) concept. After analyzing various levels of connections establishment through MAC level and respective scheduling protocols, a conclusive remark for broadcast oriented approach with data packet forwarding on geographical and temporal context being preferred. In regard of safety application known doubts, such preferable measures are required for possible connectivity but these may be exposed to security concerns. The given pseudo code with plotted graphs show impressive outputs, whether applicable for an unpredicted case is not identified. The analysis of routing protocols analyzes two location-based routing protocols. The scope of this work is limited to SIFT (SImple

Forwarding over Trajectory), and DREAM (Distance Routing Effect Algorithm for Mobility) routing mechanisms. To make a realistic approach like most of the VANET researchers prefer, a real mobility model already discussed which is used for carrying a comparative simulation study. According to similar author's view point, routing schemes for actual ad hoc networks (MANET) are not suitable for vehicular ad hoc networks (VANET). Hence the location-based schemes like SIFT and DREAM are more appropriate for such scenarios for enhanced efficiency. Further within these schemes certain issues like low-connectivity zone, spatial-awareness, and other are discussed and studied through simulations. At the end of conclusive remark, SIFT wins the game for efficiently resolving these issues. In addition, the considerations of routing in large scale are further studied.

With the help of all these major interrelated sections of mobility and routing explored in this literature review, a formal methodology be derived to consider the overall aspects of mobility and reliability with different VANET routing schemes in a pragmatic simulation environment with the help of NS2. Depending on this study, there are various routing protocols which are proposed with their suitability in MANET and VANET perspectives. After careful consideration, following are some chosen ones for the selected of city and expressway density levels of this study:

DSDV: Destination-Sequence Distance-Vector is a table driven routing protocol where every node maintains a table of information (which updates periodically or when change occurred in the network) of presence of every other node within the network. Any change in network is broadcasted to every node of the network.

AODV: Ad hoc On-demand Distance Vector an improved version of DSDV, as its name suggest, establishes the route only when demanded or required for the transmission of data. By this mean, it only updates the relevant neighboring node(s) instead of broadcasting every node of the network i.e. it does not make source routing to the entire node for the entire network.

AOMDV: Ad hoc On-demand Multipath Distance Vector, an extension of AODV with an additional feature of multipath route discovery which prevents this on-demand routing protocol to form any loop or alternative paths.

DSR: Dynamic Source Routing, maintains the source routing, in which, every neighbor maintains the entire network route from source to the destination [6-12].

## 4. IMPLEMENTATION METHODOLOGY

Vehicular Ad Hoc Network (VANET) and its related research studies are still in progression phases. The limited practical deployable options under different projects are purely simulation based before their actual implementations in the real scenarios. The list of all projects along with some related developments could be found in (VANET Projects). For this purpose, we adopted methodology for the results of this research work is based on SUMO (the helping tool for traffic mobility patterns generation for network simulator is used) and NS-2.34 simulation near to the real time packages using 802.11p protocol as shown in fig.2.

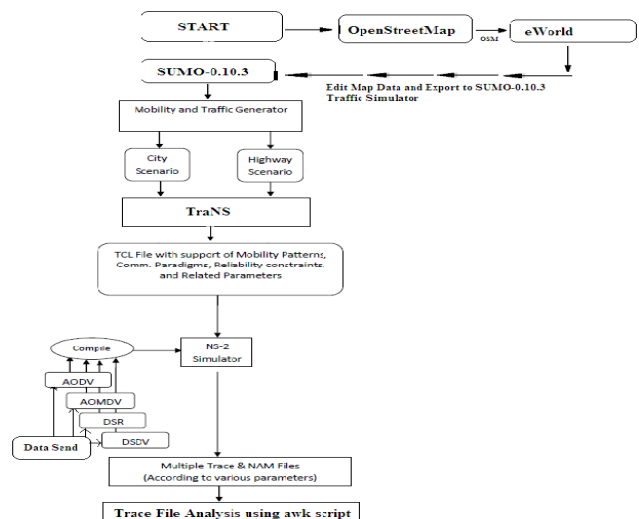

Fig.2 Flow Methodology

## 5. OBSERVATIONS

The main observation factors are related with the calculations of particular routing metrics. They identify the accumulated results from the output trace files which are generated by the simulator upon their specified inputs from mobility and traffic files. There are various routing metrics devised in different literatures to signify the importance and measuring purposes of numerous routing protocols. Realistic Traces, complete surveys along with the taxonomy of these metrics with their particular classifications. The two highly discussed metrics which are very useful in differentiating the performing trends of routing; and specially picked by similar assessments of such protocols while analyzing ns-2 traces are used for results generation in this research project. They are "Packet Delivery Ratio / Packet Delivery Fraction" and "Average End-to-End Delay".

Packet Delivery Ratio (PDR): It is the fraction of packets generated by received packets. That is, the ratios of packets received at the destination to those of the packets generated by the source. As of relative amount, the usual calculation of this system of measurement is in percentage (%) form. Higher the percentage, more privileged is the routing protocol.

Average End-to-End Delay (E2E Delay): It is the calculation of typical time taken by packet (in average packets) to cover its journey from the source end to the destination end. In other words, it covers all of the potential delays such as route discovery, buffering processes, various in-between queuing stays, etc, during the entire trip of transmission of the packet. The classical unit of this metric is millisecond (ms). For this metric, lower the time taken, more privileged the routing protocol is considered.

## 6. SIMULATION RSULTS

The navi-mumbai city model is considering the road patterns according to real scenarios. The specified regions are from the Google Maps which are transferred as input to SUMO (traffic Simulator) and then through TraNS interface the input is given to Network Simulator as shown in fig.3

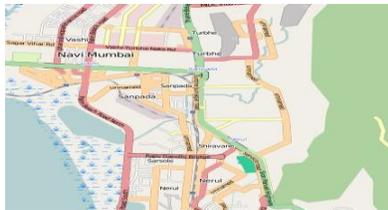
fig. 3

To study and analyze the selected routing protocol behavior while disseminating data between inter-vehicle communications, an approach of density formulation among traffic flow is used. For this reason, navi-mumbai city scene is further sub-classified on the basis of their participating vehicles in a low, medium, and high density phases.

To generate simulation instance, there are certain number of variables required to be defined within the simulation script to take action. For different densities of navi-mumbai city model, the common variables defined are shown in Table 1 below:

Table 1: Common variables in navi-mumbai city model

| Variable | Value |
|---|---|

| Simulation Time | 1000 s |
|---|---|
| Routing Protocols | AODV, AOMDV, DSR, DSDV |
| No. of Vehicles | 11, 60, 1218 |
| MAC Layer | 802.11p |
| Traffic Type | UDP |

### 6.1 Low Density Model

Simulation Parameters: The low density model along with previous common parameters comprises of 11 vehicles nodes. These nodes are deployed and arranged according to the provided patterns of mobility traces with a maximum of 8 intercommunication connections.

Simulation Results: The analyzed results from the particular trace file of the navi-mumbai city low density Vehicular Ad-hoc Network scenario are assessed on the basis of two metrics of routing protocol i.e. "*PDR / PDF*" and "*Average E2E Delay*". The outputs of their numerical calculations are placed in Table 2.

Table 2. Analyzed data of Navi-Mumbai city low density

| Navi-Mumbai City Low Density | | | | |
|---|---|---|---|---|
| | AODV | AOMDV | DSR | DSDV |
| PDR / PDF | 36.24 % | 35.58 % | 33.20 % | 34.85 % |
| E-2-E Delay | 1.59 ms | 1.57 ms | 1.49 ms | 1.50 ms |
| **Navi-Mumbai City Low Density** | | | | |
| | AODV | AOMDV | DSR | DSDV |
| Packet Loss (%) | 56.97 % | 63.34 % | 1.26 % | 64.86 % |
| NRL (%) | 0.09 % | 0.46 % | 0.11 % | 0.00 % |
| Packets Dropped | 3756 | 4176 | 2653 | 4345 |
| Packets Send | 6593 | 6593 | 4864 | 6593 |
| Packets Received | 2389 | 2293 | 1761 | 2232 |
| Routing Packets | 215 | 1060 | 192 | 0 |
| Average Throughput (Kbps) | 326.37 | 313.26 | 324.98 | 305.13 |
| Start Time (secs) | 10.01 | 10.01 | 10.01 | 10.01 |

| Stop Time (secs) | 40.00 | 40.00 | 32.21 | 39.98 |
|---|---|---|---|---|

Fig.4 Graphical Presentation of PDR for Low Density Model in Navi-Mumbai City

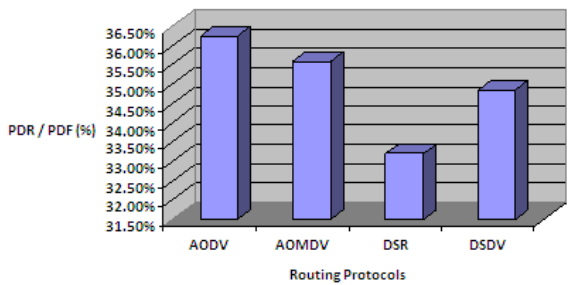

Fig.5 Graphical Presentation of E2E for Low Density Model in Navi-Mumbai City

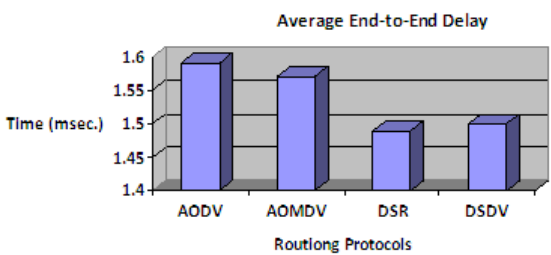

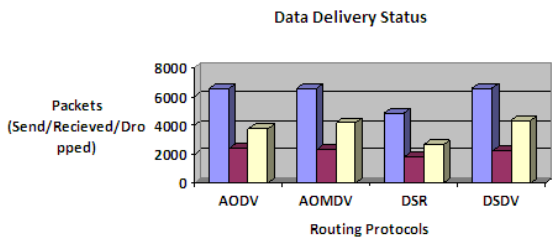

Fig.6 Graphical Presentation of Data Delivery Status for Low Density Model in Navi-Mumbai City

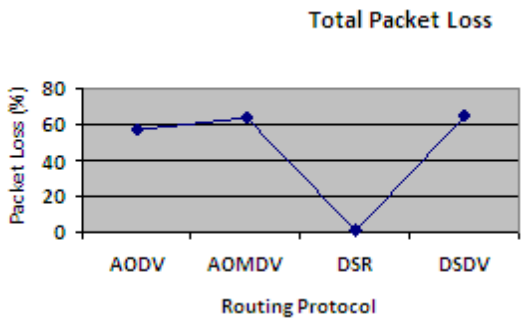

Fig.7 Graphical Presentation of Total Packet Loss for Low Density Model in Navi-Mumbai City

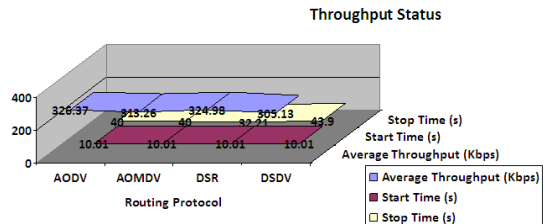

Fig.8 Graphical Presentation of Throughput Vs Routing Protocol for Low Density Model in Navi-Mumbai City

## 6.2 Medium Density Model

Simulation Parameters: The medium density model comprises of 60 vehicles as nodes and intercommunication source type of UDP with a maximum of 50 connections. Rests of the parametric values are similar to that of previous low density model declarations.

Simulation Results: The analyzed numerical results for navi-mumbai city medium density for Vehicular Ad-hoc Network scenario from the generated trace files for individual routing protocols are given in Table 3 (below).

Table 3. Analyzed data of Navi-Mumbai city medium density

| Navi-Mumbai City Medium Density | | | | |
|---|---|---|---|---|
| | AODV | AOMDV | DSR | DSDV |
| PDR / PDF | 84.35 % | 83.69 % | 14.55 % | 58.27 % |
| E-2-E Delay | 126.99 ms | 278.57 ms | 75.49 ms | 198.50 ms |

| Navi-Mumbai City Medium Density | | | | |
|---|---|---|---|---|
| | AODV | AOMDV | DSR | DSDV |
| Packet Loss (%) | 35.68 % | 36.67 % | 0 % | 40.36 % |
| NRL (%) | 4.05 % | 3.37 % | 17.83 % | 0 % |
| Packets Dropped | 7806 | 8023 | 0 | 9787 |
| Packets Send | 21879 | 21879 | 11 | 21879 |
| Packets Received | 14079 | 13935 | 6 | 12092 |
| Routing Packets | 57019 | 46982 | 107 | 0 |
| Average Throughput (Kbps) | 254.38 | 248.19 | 162.42 | 217.00 |
| Start Time | 10.00 | 10.00 | 10.00 | 10.00 |
| (secs) | | | | |
| Stop Time (secs) | 239.98 | 239.97 | 10.15 | 239.97 |

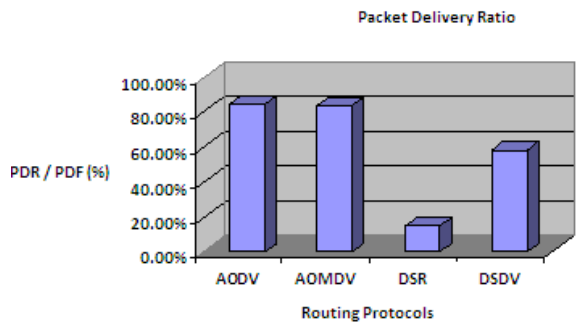

Fig.9 Graphical Presentation of PDR for Medium Density Model in Navi-Mumbai City

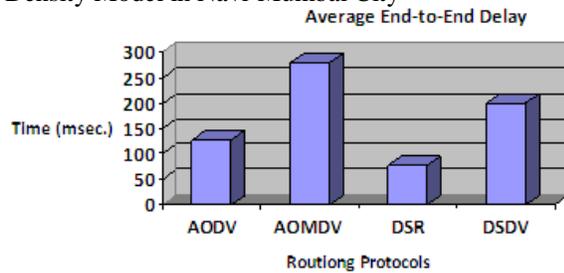

Fig.10 Graphical Presentation of E2E for Medium Density Model in Navi-Mumbai City

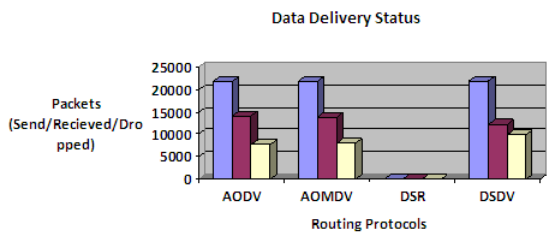

Fig.11 Graphical Presentation of Data Delivery Status for Medium Density Model in Navi-Mumbai City

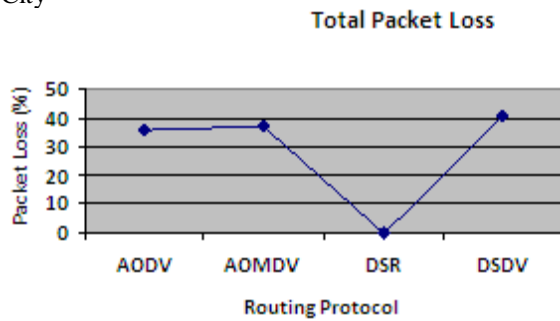

Fig.12 Graphical Presentation of Total Packet Loss for Medium Density Model in Navi-Mumbai City

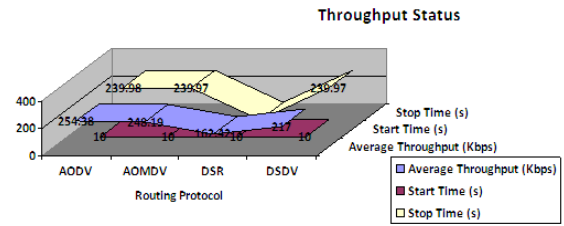

Fig.13 Graphical Presentation of Throughput Vs Routing Protocol for Medium Density Model in Navi-Mumbai City

## 6.3 High Density Model

Simulation Parameters: The high density model comprises of 1218 vehicles as nodes with all other variable values same as that of previous medium density model.

Simulation Results: The analyzed results for navi-mumbai city high density in Vehicular Ad-hoc Networks scenario with similar to previous routing metrics of "Packet Delivery Ratio" and "Average End-to-End Delay" given in Table 4.

Table 4. Analyzed data of Navi-Mumbai city high density

| Navi-Mumbai City High Density | | | | |
|---|---|---|---|---|
| | AODV | AOMDV | DSR | DSDV |
| PDR / PDF | 88.85 % | 88.89 % | 5.55 % | 38.27 % |
| E-2-E Delay | 111.17 ms | 98.36 ms | 21.49 ms | 48.50 ms |

| Navi-Mumbai City High Density | | | | |
|---|---|---|---|---|
| | AODV | AOMDV | DSR | DSDV |
| Packet Loss (%) | 77.9 % | 89.90 % | 0 % | 0.16 % |
| NRL (%) | 16.72 % | 89.46 % | 2.67 % | 0 % |
| Packets Dropped | 2341 | 1095 | 0 | 309 |
| Packets Send | 3003 | 1218 | 20 | 3654 |
| Packets Received | 160 | 125 | 18 | 404 |
| Routing Packets | 2676 | 1180 | 150 | 0 |
| Average Throughput (Kbps) | 5.95 | 324.40 | 2.12 | 16.55 |
| Start Time (secs) | 10.00 | 10.00 | 10.00 | 10.00 |
| Stop Time (secs) | 125.70 | 11.58 | 8.50 | 126.30 |

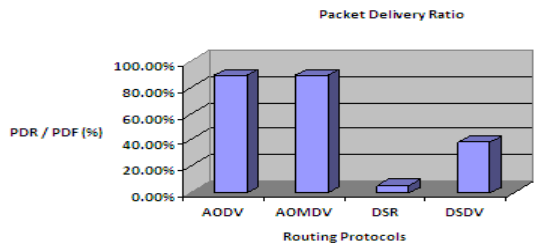

Fig.14 Graphical Presentation of PDR for High Density Model in Navi-Mumbai City

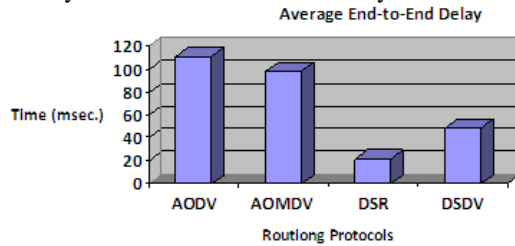

Fig.15 Graphical Presentation of E2E for High Density Model in Navi-Mumbai City

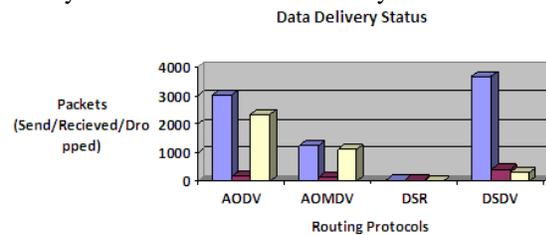

Fig.16 Graphical Presentation of Data Delivery Status for High Density Model in Navi-Mumbai City

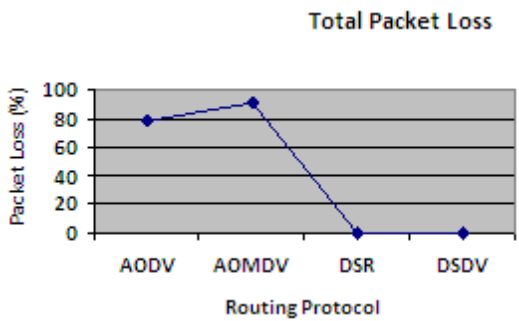

Fig.17 Graphical Presentation of Total Packet Loss for High Density Model in Navi-Mumbai City

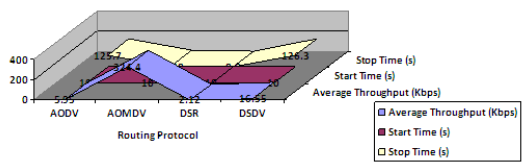

Fig.18 Graphical Presentation of Throughput Vs Routing Protocol for High Density Model in Navi-Mumbai City

## 7. CONCLUSION

In this report I would like to conclude that I have taken the real city scenario and tried to run 1200 vehicles in an hour and 20,000 to 22,000 in a day and check the results using simulators. We have used different simulators to obtain the results and send the data between inter-vehicle properly according to the incident occurred on road which helps in avoiding accidents and traffic jam and also we have tried to analysis the best routing protocol which will help us to send our data and give better performance. So the results obtain DSR routing protocol using DSRC/IEEE 802.11p gives use better performance results to send our own message according to the event occurred. We have also checked the results on national highway where in a day 24,000 vehicles pass through the express highway and V2V communication using DSRC/IEEE 802.11p and also using DSR routing protocol gives better performance results as compared to other routing protocols. In future we may increase the strength of vehicles in city and on national/express highway and check the results and also provide other information to the vehicle operators (i.e. nearest petrol pumps, hotels etc.)

.